\documentclass[aps,prl,twocolumn,groupedaddress,showpacs]{revtex4}

\usepackage{graphicx}
\usepackage{dcolumn}
\usepackage{bm}

\begin{document}

\title{Photon generation in an electromagnetic cavity with a time-dependent boundary}

\author{C.M.~Wilson$^1$, T.~Duty$^{1,2}$, M.~Sandberg$^1$, F.~Persson$^1$, V.~Shumeiko$^1$, P.~Delsing$^1$}
\affiliation{$^1$Microtechnology and Nanoscience, Chalmers University of
Technology, S-41296, G\"{o}teborg, Sweden.}
\affiliation{$^2$School of Mathematics and Physics, University of Queensland;
St. Lucia, QLD 4072 Australia.}
\date{\today}

\begin{abstract}
\noindent We report the observation of photon generation in a microwave cavity with a time-dependent boundary condition.  Our system is a microfabricated quarter-wave coplanar waveguide cavity. The electrical length of the cavity is varied using the tunable inductance of a superconducting quantum interference device.  It is measured in the quantum regime, where the temperature is significantly less than the resonance frequency ($\sim 5$ GHz).  When the length is modulated at approximately twice the static resonance frequency, spontaneous oscillations of the cavity field are observed.  Time-resolved measurements of the dynamical state of the cavity show multiple stable states. The behavior is well described by theory.  Connections to the dynamical Casimir effect are discussed.
\end{abstract}


\maketitle


Photons, in contrast to electrons, do not directly interact with each other.  However, effective interactions can be induced when photons interact with a nonlinear media.  In many cases, these effective interactions result in so-called parametric processes which are very important in widespread technological applications and also in fundamental studies of quantum electrodynamics \cite{Pan:1997p2204,Wu:1986p2366}.  In the optical regime, the effective nonlinearities are generally bulk properties of materials or plasmas.  In the rf and microwave regime, they are often created by lumped-element electrical devices \cite{Yurke:1989p2575}.  These types of systems have been extensively studied both theoretical and experimentally.

Alternatively, it has been proposed that parametric processes can be observed in a system that is essentially linear, but where a boundary condition of the electromagnetic field can be changed rapidly in time.  For instance, this is the central theoretical problem in the field of the Dynamical Casimir effect (DCE) \cite{Moore:1970,Fulling:1976p2777}.  One striking prediction of the DCE is that real photons can be generated out of the vacuum by the changing boundary condition.  The experimental  system often imagined in describing this vacuum DCE is a cavity with a moving mirror.  Estimates suggest this is technically a very challenging route to pursue, however.  Therefore, it has been suggested that the observation of photon generation starting from a thermal field, which we will call the thermal DCE, would be much easier, while still demonstrating many of the essential features of the vacuum DCE \cite{Plunien:2000p2024}.  Alternatively, it has recently been proposed that the same physics can be studied by instead modulating the \textit{electrical} length of a cavity \cite{Johansson:2009p2003}.

Regardless of the source of the interactions, the quantum behavior of parametric systems has been of great interest.  For example, the quantum dynamics of parametric oscillators (PO) has been described theoretically in several different contexts \cite{Dykman:1998p1044,MilburnHolmes}. In fact, the DCE can be mapped to the quantum version of the PO in special cases \cite{Dodonov:1995p1991}.  A driven PO can exist in a number of stable dynamical states.  In a PO with negligible loss, it is predicted that quantum tunneling between the states is possible.  This coherent coupling between the states is then predicted to lead to the formation of so-called Schr\"odinger cat states.  In POs with more loss, it is predicted that quantum tunneling will always be hidden by a distinct process know as quantum activation \cite{Marthaler:2006p995}.  Quantum activation is an incoherent process, driven by relaxation (spontaneous emission), that nonetheless leads to "over-the-barrier" switching of the dynamical state, even at zero temperature.  None of these quantum predictions have been conclusively observed in experiment.

\begin{figure}[t]
\includegraphics[width = 0.92\columnwidth]{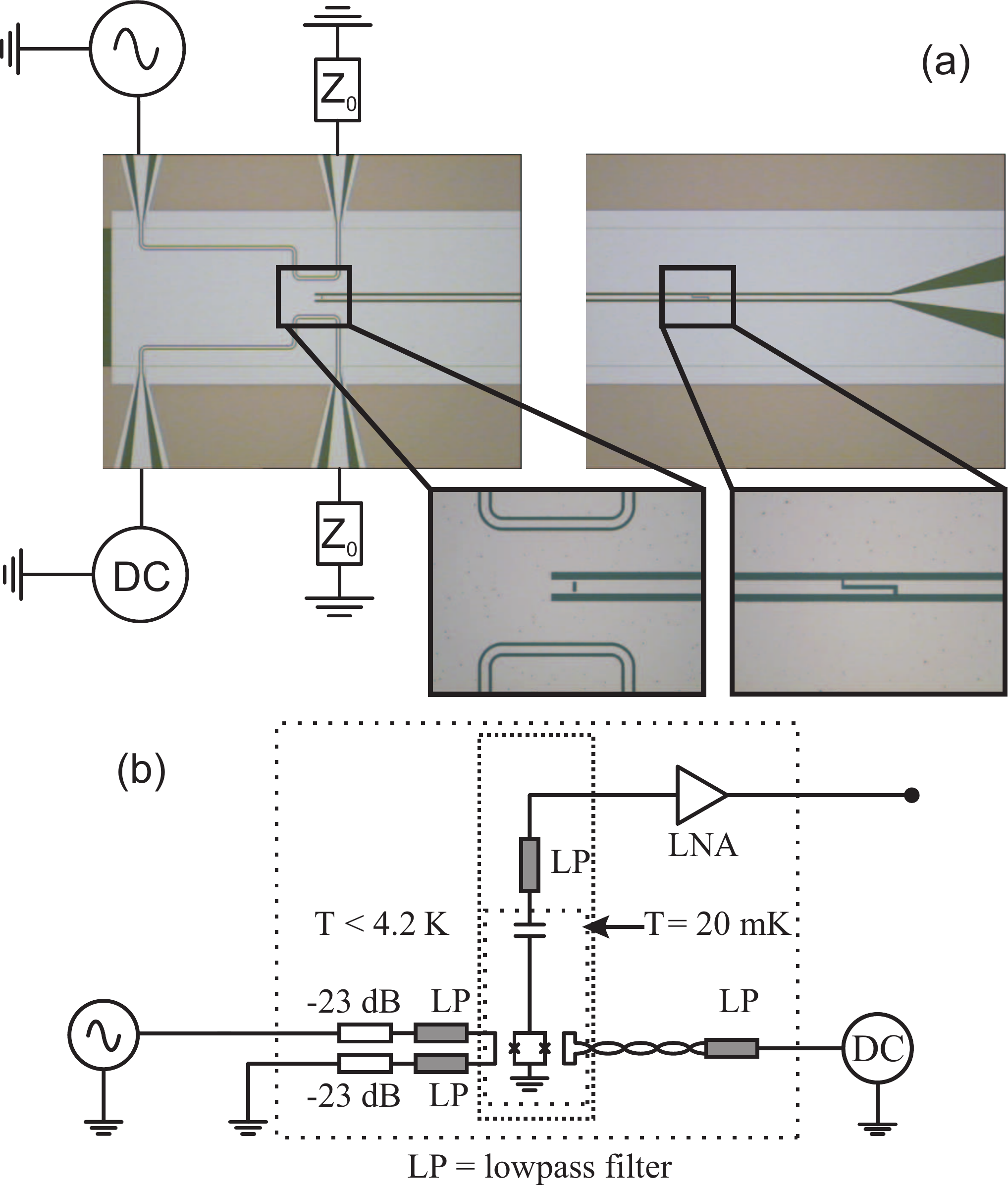}
\caption{a) A micrograph of a quarter-wavelength cavity. The cavity is probed through a small coupling capacitance while the other end is terminated to ground through a SQUID. Changing the external magnetic field through the SQUID loop changes the boundary condition of the cavity. b) Simplified block diagram. The cavity is measured using a circulator and a cold amplifier. The magnetic field is applied via on-chip control lines, one for high frequency and one for DC.}
\label{Schematic}
 \end{figure}

To this end, superconducting systems are interesting for several reasons: Josephson junctions can be used to make the resonators frequency tunable over a wide range; various nonlinearities can be easily designed at a desired strength; and all of this can be done with very little dissipation. In addition, superconducting resonators easily work at microwave frequencies, implying that operation in the quantum regime is possible using standard cryogenic techniques.   Parametric devices based on Josephson junctions were pioneered in the 80s by Yurke \textit{et al.} \cite{Yurke}, but have recently regained interest  \cite{Tholen:2007p4165, CastellanosBeltran:2008p4574, Yamamoto:2008p1126, Sandberg:2008p072102,PalaciosLaloy:2008p3486,BuksEPL09}.

\begin{figure*}[t]
\includegraphics[width=2\columnwidth]{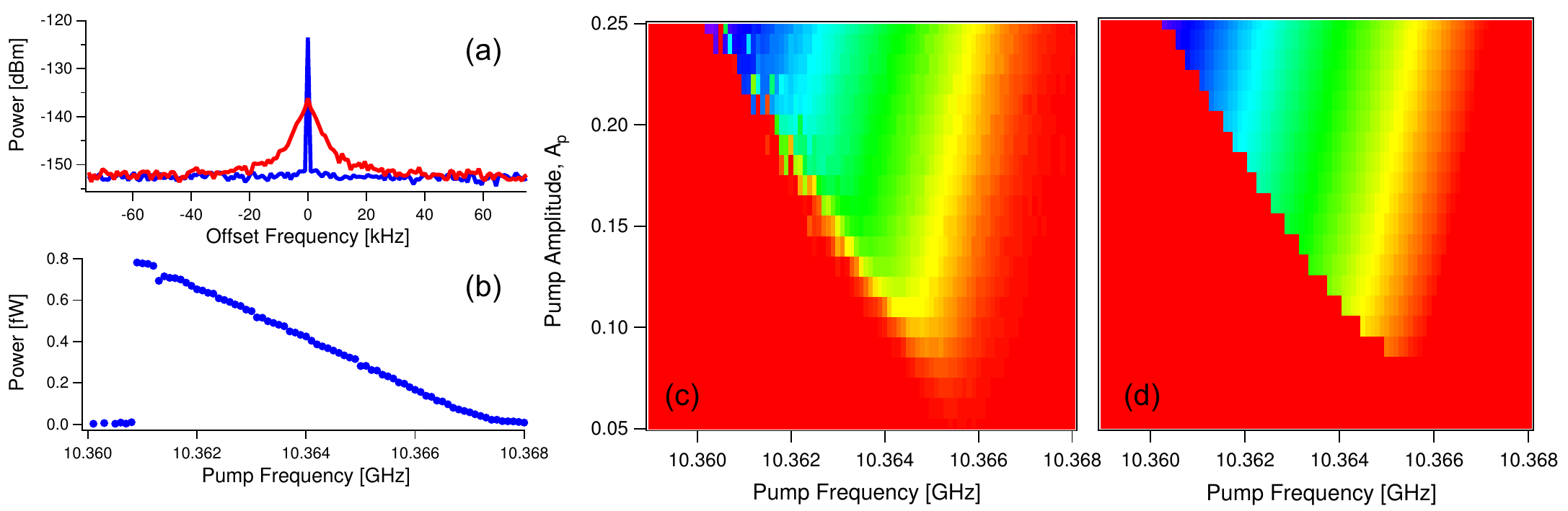}
\caption{(Color Online) The output of the pumped cavity. a) Two output power spectra referred to the cavity output, for different working points.  The offset frequency is centered at $f_p/2$.  The widths of the peaks are the switching rates between different states (see Fig. \ref{Metapotential}).  b) Integrated output power as a function of $f_p$ for $A_p = 0.23$.  The amplifier noise power has been subtracted. The predicted linear dependence on detuning from the bifurcation point is clear.  c) Integrated output power as a function of $A_p$ and $f_p$. d) Theoretical fit from solving equation\,\ref{eq:diffeq}. The color scales are the same in both plots.}
\label{amplitude}
\end{figure*}

In this work, we have studied tunable, high-Q superconducting cavities.  They are  quarter-wavelength coplanar waveguide cavities fabricated on-chip (Fig.\,\ref{Schematic}).  The cavities are made tunable by incorporating a Superconducting QUantum Interference Device (SQUID), which is used as a parametric inductance. The cavities are cooled such that $k_BT \ll\hbar \omega_0$, $\omega_0$ being the resonance frequency.  This means that the average photon number in the cavity is well below one, \textit{i.e.}, the cavity is in the quantum regime. When we strongly drive the magnetic flux through the SQUID at approximately twice the resonance frequency, we observe the generation of microwave photons in the cavity (Fig. \ref{amplitude}a).

Our cavities are spatially extended systems described by a wave equation.  The modulated SQUID imposes a time-dependent, nonlinear boundary condition on the wave equation.  For a static but nonlinear boundary condition, an exact, analytical solution is already not possible, although, approximate solutions can be found in the form of coupled oscillators \cite{Wallquist:2006p2000}. Treating time-dependent boundary conditions is one of the essential aspects of the Dynamical Casimir problem.  For a general time-dependent boundary condition, the system can be reduced to an effective set of coupled modes where both the mode frequencies and the couplings are time dependent   \cite{Dodonov:2001p1149,Schutzhold:1998p1162}.  If the cavity is designed such that the static mode frequencies are not multiples of each other, then we can find an approximate solution when the modulation frequency is twice the static resonance frequency of one of the modes.  In this case, the dynamics of the isolated, pumped mode can be reduced to those of a PO \cite{Dodonov:1995p1991}.  Starting from a pure vacuum state in the cavity, the appearance of parametric oscillations in this mode is then one example of the vacuum DCE.  

Some care must therefore be taken in describing the dynamics of the system, taking into account the difficulties and subtleties mentioned above.  That said, we do find experimentally that the dynamics of the system can be described in a reasonably quantitative fashion by (\ref{eq:diffeq}) with the appropriate set of effective parameters.  The dynamical variable is the canonical flux of the cavity $\Phi = \Phi_0\varphi/2\pi = \int_{-\infty}^t V dt'$ where $V$ is the voltage and $\Phi_0$ is the flux quantum.

A PO can be described by the differential equation \cite{Dykman:1998p1044}:
\begin{equation}
\partial_{tt} \varphi +2 \Gamma \partial_t \varphi +  \left( \omega_0^2 + F\cos \left( \omega_p t\right) \right)\varphi+\gamma \varphi^3=\xi(t),
\label{eq:diffeq}
\end{equation}
\noindent where $F$ is the amplitude of the frequency modulation, $2\Gamma=\omega_0/Q$ is the linewidth of the resonance and describes the damping of the system, and $\gamma$ represents the dominant nonlinearity of the system, the so-called Duffing term. $\xi(t)$ represents a mean-zero noise term that leads to activated switching.  

We derive an expression for $F$ by linearizing the tuning curve of the cavity $\omega_0^2 = \omega_b^2(1+r(x))^{-2}$ with respect to the external flux $x = \Phi_B/\Phi_0$ \cite{Sandberg:2008p072102}.  Here, $r(x) = L_s(x)/dL_l$ where $L_s$ is the SQUID inductance, $d$ is the physical length of the cavity, and $L_l$ is its inductance per unit length. We find $F = 2 \pi \tan(\pi x_{dc})(\omega_0^3/\omega_b)r x_{ac}$ where the subscripts dc and ac distinguish between the static bias point and ac pump amplitude. In our system, $\gamma$ arises from the coupling of the current in the excited cavity to the SQUID.  Following \cite{Wallquist:2006p2000}, we find $\gamma = -\omega_0^2 \delta^3/\pi$ with $\delta = -(\pi/2)r/(1+r)$.



Once the system has been put in the form of (\ref{eq:diffeq}), its behavior can be understood in terms of the ÒslowÓ quadrature variables, $q_1$ and $q_2$, where $\varphi(t)=q_1(t) \cos(\omega_p t) - q_2(t) \sin(\omega_p t).$ In the rotating frame, the dynamics of $q_1$ and $q_2$ are determined by the metapotential
\begin{equation}
g(q_1,q_2)=\frac{\Omega}{2}\left(q_1^2+q_2^2 \right)+\frac{\zeta}{2} \left(q_2^2-q_1^2 \right)+\frac{\beta}{4} \left(q_1^2+q_2^2 \right)^2
\label{eq:metapotential}
\end{equation}
\noindent where $\Omega=\frac{1}{\Gamma}(\frac{\omega_p}{2}-\omega_0$) is the normalized detuning, $\zeta=F/2\Gamma\omega_p$ is the normalized drive strength, and $\beta = 3\gamma/4\Gamma\omega_p$ is the normalized nonlinearity. Below the pump threshold value of $\zeta = 1$, this potential has only one minimum centered at the origin. Thus the system does not oscillate. For small detunings, as the threshold is crossed, two symmetric minima develop, yielding two stable, oscillating states phase shifted by $180^{\circ}$. As the pump is blue detuned, we reach a bifurcation point where the two stable states merge into a single "quiet" state at the origin. If the pump is instead red detuned, another bifurcation point is reached where a metastable state develops at the origin. The system then has three states, the two $\pi$-shifted oscillating states and a quiet state.  

The cavity and SQUID are fabricated in aluminum using the standard double-angle, shadow evaporation technique. We have previously shown that the frequency of such a cavity can be changed much faster than the lifetime of the photons in the cavity \cite{Sandberg:2008p072102}. The samples were mounted in a magnetically shielded sample holder in a dilution refrigerator with a base temperature of approximately 20 mK. The measurements of the cavities were done using a cryogenic amplifier  at 4.2 K that had a nominal noise temperature of 4 K. The sample was connected to the amplifier via a circulator mounted on the mixing chamber. The signal was further amplified at room temperature. The resonance frequencies and the Q-values of the samples were characterized by coupling a heavily attenuated probe signal to the cavity via the circulator.  For the photon generation experiments, no signal was applied to the cavity but a flux pumping signal at approximately twice the resonance frequency was applied to the fast flux line.  The output of the pumped cavity was then recorded using a vector digitizer.
 
\begin{figure*}[t]
\includegraphics[width=2\columnwidth]{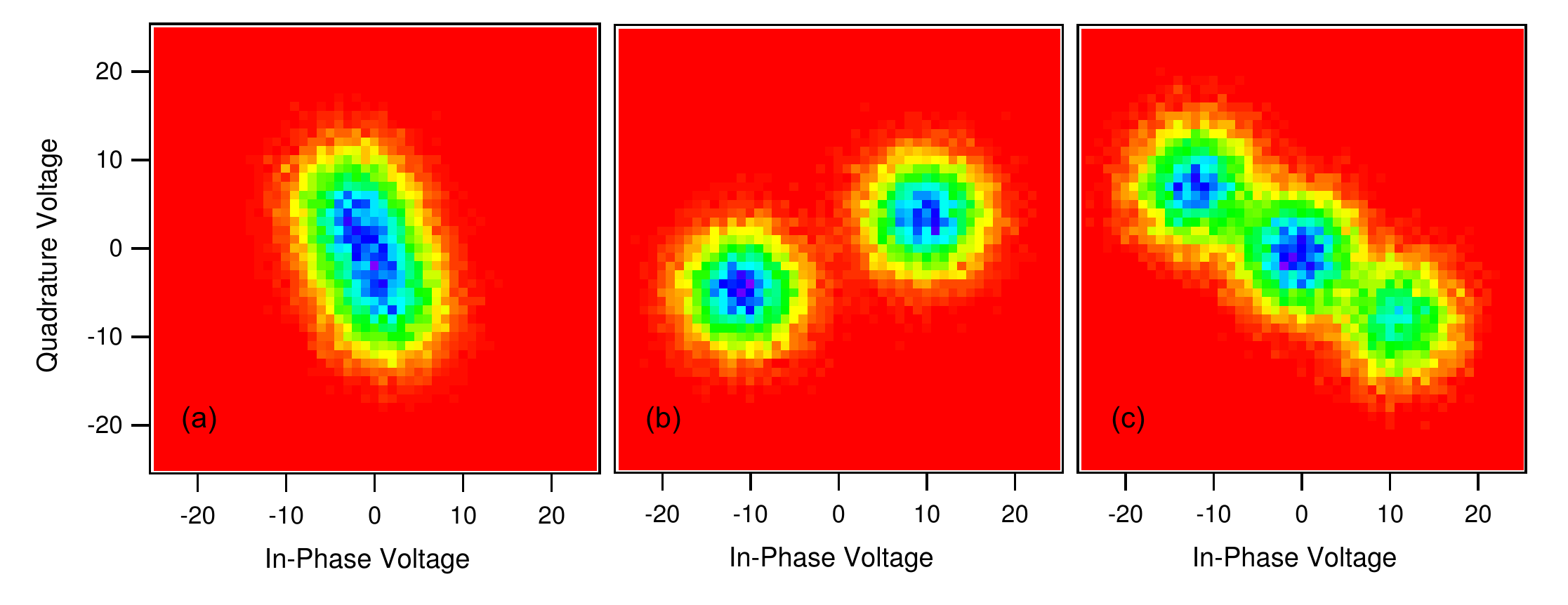}
\caption{(Color Online) Exploring the metapotential.  2D histograms of measured in-phase, $q_1$, and quadrature, $q_2$, voltages for three different working points.  (a) Just below threshold, the metapotential softens in one direction and we observe a noise ellipse.  (b) Above threshold, the system bifurcates, exhibiting two finite-amplitude oscillating states.  These states have equal amplitude, but are shifted by $180^{\circ}$ in phase.  (c) When the pump is red detuned from the cavity, a quiet state can coexist with the oscillating states, in agreement with theory.  The phase rotations between the three images are instrumental artifacts.}
\label{Metapotential}
\end{figure*}

We have studied the pumped cavities as a function of both pump amplitude and frequency. In Fig.\,\ref{amplitude}, we show the output power of a pumped cavity with $\omega_0/2 \pi = 5.18$ GHz and $Q = 8900$. The pump amplitude, $A_p$, is scaled such that $20\log(A_p)$ is the power at the microwave generator in dBm.  In Fig.\,\ref{amplitude}\,d, we show the steady state solution to (\ref{eq:diffeq}), namely $\langle \varphi^2 \rangle = (\sqrt{\zeta^2-1}-\Omega)/2\beta$.  In the range where this solution is imaginary or negative, we take the solution $\langle \varphi^2 \rangle = 0$.  To scale the y-axis of the theory, we define $F = F_A A_p$ and calculate a value of $F_A = 9.9 \times 10^{18} s^{-2}$ using a dc calibration of the pump flux coupling.  To scale the z-axis (color scale), we calculate $\gamma = 1.9 \times 10^{17} s^{-2}$ and estimate the total gain of our amplifier chain, which is nominally $G = 75$ dB. To fit the data, we allow $F_A$ and $G$ to vary.  We find a best fit for $F_A = 3.0 \times 10^{18}$ and $G = 73$ dB.  We note that since the power is proportional to $G/\gamma$, these parameters are not independently constrained. It is not surprising that we would find a different value for $F_A$ since the cable loss and flux mode structure should be different for the dc calibration and the $\sim 10$ GHz pump flux.  Keeping that in mind, we conclude that the agreement is good.

The data is clearly asymmetric with respect to frequency.  This is explained in the following way.  At the blue detuned bifurcation point $\Omega_B^{+} = +(\zeta^2 -1)^{1/2}$, the oscillating states vanish and only the quiet state is stable.  In contrast, at the red detuned $\Omega_B^{-} = -(\zeta^2 -1)^{1/2}$, the quiet state emerges, but is only metastable, with a occupation probability that is exponentially small compared to the oscillating states.  The oscillations only stop when the occupation probability of the quiet state becomes significant.  In \cite{Dykman:1998p1044}, it is estimated that this happens for $\Omega \approx -4 \zeta$.  We therefore introduce by hand a cutoff on the red detuned side, although we find that $\Omega = -3.4 \zeta$ agrees better with the data.



We can visualize the dynamics of the system by making histograms of the measured quadrature pairs $(q_1,q_2)$.  The maxima of the histograms then correspond to the stable points of the metapotential.  In this way, we can map out the metapotential (\ref{eq:metapotential}).  In Fig.\,\ref{Metapotential}, we plot histograms of $(q_1,q_2)$ sampled at 1 MHz.  In agreement with theory, we find that the system has three qualitatively different conditions: 1) one stable state where the magnitude is zero, 2) two stable states symmetric about zero, 3) three states which combine 1) and 2).  We can also clearly see the softening of the metapotential just before the system bifurcates. 

We observe switching between the different states of the system. The switching may be caused by a variety of mechanisms including thermal activation, quantum activation, or quantum tunneling \cite{Marthaler:2006p995,Serban:2007p21500}.  A detailed analysis of the observed rates including their dependence on various parameters is needed to distinguish these mechanisms.  This will be the topic of a future paper.  

We thus find a good agreement between the response of a cavity with a time-dependent boundary condition and the theory of a PO.  This experimentally confirms one of the fundamental predictions of the DCE literature \cite{Dodonov:1995p1991}.  Furthermore, the quantitative agreement demonstrates that the source of photon generation is in fact the time-dependent boundary and not some other mechanism.  In conclusion, we can interpret this work as an observation of the thermal DCE.


In fact, this experiment has been performed in the quantum limit. Thus it is fair to ask whether we can interpret these results in terms of the vacuum DCE.  This is essentially a question of whether quantum fluctuations or thermal fluctuations initiated the cavity oscillations. We cannot conclusively distinguish between the classical and the quantum result, since we only measure the steady state oscillations which are insensitive to the initial conditions.  In the future, it may be possible to distinguish the quantum and classical result by observing the system's transient response.  Alternatively, the steady state solutions are different if the cavity is removed and the SQUID is left to modulate the boundary condition of an open transmission line \cite{Johansson:2009p2003}.

We would like to thank Mark Dykman, G\"oran Johansson, and the Quantum Device Physics and Applied Quantum Physics groups for useful discussions.  The samples were made at the nanofabrication laboratory at Chalmers. The work was supported by the Swedish VR,  the Wallenberg foundation, and by the European Union through the ERC and the projects EuroSQIP and SCOPE. 

\bibstyle{apsrev}

\end{document}